# Glass-like kinetic arrest of first-order structural phase transition in $La_xMnO_{3\pm\delta}$ (x= 1 & 0.9)

Aga Shahee, Dhirendra Kumar and N. P. Lalla*

*UGC-DAE Consortium for Scientific Research, University campus, Khandwa road Indore, India- 452017*

## Abstract

We report here the occurrence of glass-like kinetic-arrest of first-order phase transition of R-3c to Pbnm in a supercooled $La_xMnO_{3\pm\delta}$ (X= 1 & 0.9 and $\delta$>0.125) below its insulator-metal transition employing low-temperature transmission electron microscopy (TEM) and x-ray diffraction. No transformation was observed even in a ~12 hrs aged $La_xMnO_{3\pm\delta}$ at 98K until unless it was triggered insitu using e-beam irradiation in TEM. A systematic triggering at different temperatures has revealed the occurrence of kinetic-arrest situation similar to stable glass-formers. This has been attributed to the suppression of Johan-Teller distortion by double-exchange.

*Corresponding author: N. P. Lalla (nplalla@csr.res.in)

The critical slowing down of the dynamics of molecules with decreasing temperature results in "kinetic-arrest", ceasing the homogeneous nucleation of the supercooled liquid into crystalline solid and consequently resulting into a liquid to "Glass-transition"[1]. Although the phenomenon of glass-transition is known since long its physics is yet not well understood [2, 3]. The term "glass" is often loosely taken as synonyms of the "structural disorder" but the widely accepted definition is a liquid in which the molecular motion has under gone "kinetic-arrest". This term has mostly been discussed in relation to disorder-disorder (liquid to glass) phase transitions (PT), but now it has been realized in order-order (OO) PTs too. Chattopadhyay et al. have shown [4] that the pseudobinary alloy $Ce(Fe_{0.96}Ru_{0.04})$ under goes kinetic-arrest of the first-order ferromagnetic (FM) to antiferromagnetic (AFM) transition at low-temperatures (LT) when the applied field is above a critical value. Kinetic-arrest in OO type PT has more frequently been reported in perovskite ($ABO_3$) manganites[5-7] and multiferroics[8]. Using magnetic force microscopic imaging Weida et al.[5] reported that $La_{5/8-y}Pr_yCa_{3/8}MnO_3$ ($y$ = 0.4) goes to a magnetic-glass state at LT due to cooperative, dynamic freezing (kinetic-arrest) of the first-order AFM to FM transition. Banerjee et al [6] and Chaddah et al [7] have reported the magnetic-glass state in half-doped manganites $Pr0.5Ca_{0.5}Mn_{0.975}Al_{0.025}O_3$ and $La_{0.5}Ca_{0.5}MnO_3$ respectively. They have demonstrated the occurrence of field-induced dearrest of stable FM phase out of kinetically arrested metastable AFM phase [6] and kinetic-arrest of FM phase in a AFM phase [7] at LT.

The occurrence of kinetic-arrest in OO first-order phase transitions (FOPT) transitions appears to generalize its fundamental role played in "Glass-transition" to other FOPT too. The entropy of the high-temperature liquid phase is preserved in the form of disorder of the atomic arrangement in the glassy phase. Thus the study of kinetic-arrest in a verity of OO transitions becomes important from basic interest point of view that in what form the entropy is preserved in these arrested phases. Thus it will be far more intriguing if such a kinetic arrest is realized in an OO structural PT.

Manganites [9] pose spin, charge, orbital and lattice degrees of freedom [10], which result in correlated magnetic, electronic and structural PTs[11-15]. These correlations basically originate from phase-separation [16,17] caused by chemical disorder at A and B sites[18] and the coexistence of localized and band-like wave-functions of $e_g$-electrons [19]. Keeping in view the close association of the kinetics of the phase transition and the accommodation strain [5], the kinetic-arrest of OO structural transition is quite likely to occur in manganites itself.

Here we report the occurrence of glass-like kinetic-arrest of R-3c (or I2/a) to Pbnm FOPT at LT in oxygen excess and self-doped $La_xMnO_{3\pm\delta}$ prepared through solid-state rout (calcined at 1100$^o$C and sintering at 1250$^o$C for 24 hrs). We also show that competing double-exchange (DE) [20, 21] and JT-distortion (JTD) causes the kinetic-arrest of R-3c (or I2/a) to Pbnm FOPT at LT. LT TEM and XRD have been used for this study. No signature of any phase-transformation, i.e. the occurrence of new spots, could be seen in the <241> zone pattern recorded immediately after the sample temperature got stabilized at 98K after being cooled with a rate of ~10K/minute, Fig.1a. Keeping in view the possibility of slow kinetics, the sample temperature was maintained at 98K for ~12hrs and SAD was again recorded exactly from the same area, see Fig.1b. Even then there was no signature of any transformation. A fine probe of ~30nm was then made and focused on the region of interest for few seconds. A small black contrasted region (nucleus) of ~80nm was found to appear at the trigger point. Microstructure and SAD from that region were immediately recorded, see Fig.1c,d. The occurrence of new spots (marked by arrows) in Figs.1d show that e-beam triggering has initiated structural PT of R-3c.Once nucleated at 98K the Pnma region was found to grow on its own without any further beam irradiation and give rise sharp and intense diffraction spots, as indicated by thick arrows in Fig.2b. The occurrence of a new HOLZ ring, with smaller radius, in CBED shown in Fig.2c corresponds to ~7.78 Å, which shows doubling of the periodicity along the beam direction. The pattern in Fig.2b corresponds to [001] zone SAD pattern of Pbnm phase. Thus e-beam irradiation trigged the transformation of rhombohedral R-3c phase with cell parameters of a=5.53 and c=13.37 Å to orthorhombic Pbnm phase with cell parameters of a=5.4 Å, b=5.5 Å and c=7.78 Å. The large difference of ~0.1 Å in 'a' and 'b' parameters indicates that the structure of the transformed Pbnm phase is JT-distorted O' type[22-26], which is insulating. The presence of weak spots as indicated by thin arrows shows the occurrence of charge-orbital ordering. The Pbnm phase, appeared due to e-beam triggering at LT, was found to transform back to R-3c on heating, see Fig.2d.Transformation starts at ~219K and finishes at ~260K showing distinct phase coexistences of Pbnm and R-3c confirming to the FOPT, see intensity-vs-temperature plot in Fig.3b. The FM to paramagnetic PT across the MI-transition was insitu monitored in TEM through e-beam deflection caused by the Lorentz-force exerted by the sample magnetization induced by ~2.2T field of the TEM objective. Fig.3b (left axis) presents the plot of the beam deflection as a function of temperature. It can be seen that the deflection amplitude, i.e. the net-magnetization, of the sample becomes zero at ~255K, which matches well with the MI transition temperature $T_c$ in the R-T measurement. Transformation of R-3c FM-metallic and O'-type AFM-insulating phases back to same R-3c PM-insulating phase at the same temperature indicates that their origin is interconnected. LT XRD also did not show any PT down to 79K even after ~12 hrs aging. The XRD profiles corresponding to 79K were successfully refined [22] using R-3c.The temperature dependent XRD data does show distinct broadening of the peak with R-3c (hexagonal lattice) index (220). The (220) peak broadening increases with decreasing temperature down to ~235 K and below ~235K it again gains its sharpness as well as peak height, see inset of Fig.4.The broadening is due to initiation of distortion of R-3c to the lower symmetry phase I2/a, which gives rise to splitting of R-3c peak with index (220) into I2/a peaks with indices (-422) and (040).R-3c ($a^- a^- a^-$) and I2/a ($b^- a^- a^-$) structures differ only by an unequal anti-phase Glazer's tilt [27] about any one cartesian-axis. This difference in the tilt may arise due to distortion of the $MnO_6$ octahedra under going cooperative JTD at LT. The observed distortion appears to be a premonition of phase-transformation of R-3c to orbitally-ordered Pbnm phase [22, 23]. The continued distortion would have caused the PT. But since it gets intercepted by other competing interaction it gets suppressed reversing its direction back to the undistorted R-3c.

The observed e-beam triggered PT at LT appears to be a case of dearrest (glass-like devitrification) of the stable Pbnm phase out of kinetically arrested supercooled R-3c phase. During a FOPT when an equilibrium phase is cooled towards super-cooling spinodal ($T^*$) [28] any energy fluctuation ~ $\Delta E_b$ (energy barrier) will trigger the transformation, with probability increases exponentially as T approaches $T^*$. In the absence of energy-fluctuation an equilibrium phase can be supercooled to $T^*$ below which the supercooled phase spontaneously transforms to another equilibrium phase. If there happens to be an intercepting phenomenon above $T^*$, which some how or other inhibits the change of the order-parameter, then a characteristic temperature $T_k$ ($T_g$) exists above $T^*$, below which the expected FOPT gets "kinetically arrested" like in the case of "glass-transition".

To probe the existence of supercooling spinodal $T^*$ we triggered the transformation using at various temperatures below 260K during cooling. No transformation was seen until the sample temperature reached to 206K. Fig.5a shows an electron micrograph of nuclei, which were obtained at 206K, 190K, 178K and 151K, each after interval of ~20 minutes. The occurrence of an onset temperature, 206K and, although small but increasing size of the transformed region at LT, indicate that there may exists a $T^*$. As T is approaching $T^*$ the probability of nucleation (i.e. the size of the transformed region) of Pbnm phase after triggering, is increasing. It was seen, Fig.5a, that these nucleated regions did not grow even after an hour, where as the nucleus obtained by triggering at ~101K grew to a much larger region only in few tens of minutes, see the electron micrographs in Figs.5b-d.The absence of growth of nuclei above 151K may either be due to that T and $T^*$ are still not close enough to enhance the fluctuation induced transformation or the growth has been taken over by some phenomenon causing kinetic-arrest of the phase transition with characteristic temperature $T_k$ such that $T^* < T_k$ . Once $T^* < T_k$ like situation occurs $T^*$ looses its significance. Therefore to confirm the occurrence of kinetic-arrest, search of the existence (effectiveness) of $T^*$ was carried out. Since in manganites the transport properties sensitively depend on the structure [29], to explore the existence of $T^*$ we continuously monitored R-T variation in heating cooling cycles down to 4K in temperature steps of 1K. Down to 4K which nearly all structural phase-transformation in solids are expected to be complete, see Fig.6. Unlike the hysteresis observed in the R-T variation across the R-3c to Pbnm PT in $La_{1-x}Sr_xMnO_3$ [29], we did not observe any hysteresis in the wide temperature range of 4K to 300K for $La_xMnO_{3\pm\delta}$. Absence of hysteresis directly implies that for $La_xMnO_{3\pm\delta}$ $T^*$ has become insignificant (ineffective) and R-3c phase under goes supercooling and the kinetics required for R-3c to Pbnm phase transition is some how getting arrested below $T_k > T^*$. This is analogous to the situation usually seen in stable glass-formers where a characteristic temperature $T_g$ (or $T_k$) exists above $T^*$. In the present case it so appears that octahedral tilt adjustment necessarily required for of R-3c to Pbnm phase transition remains kinetically arrested.

The XRD results shown in Fig.4, which depict increasing and decreasing distortion in R-3c phase across ~235K, indicate that some competing interactions may be responsible for the observed kinetic-arrest of R-3c to Pbnm phase-transformation. Similar increasing decreasing distortion has also been seen in $La_{0.88}MnO_3$ through neutron scattering at LT and has been attributed to the excitation of JT- distortion modes $Q_2$ & $Q_3$ in $MnO_6$ octahedra [20]. From FWHM -vs- T and R-T plots in Fig.6a,b it is quite obvious that the distortion in the R-3c lattice starts in the vicinity of the commencement of MI transition. As the rate of initial drop in the resistance starts decreasing, i.e. as the ferromagnetism reaches its saturation, the distortion also starts decreasing. This correlation indicates that there is a competition between ferromagnetism i.e. the DE and the distortion of the R-3c lattice i.e. JT-distortion, in which DE dominates. The JT-distortion and DE are anti-correlated. Their occurrence in the vicinity of same temperature indicates the evolution of phase-separation. It has been shown [19] that in mixed-valent manganites localized 'l' and band-like 'b' characters of $e_g$-electrons coexist and at LT 'l' and 'b' sates hybridize making the JT-distortion dynamic and consequently suppressed[30]. Keeping in view the increase and decrease in the lattice distortion around $T_c$,Fig.6a,b, and the transformation of the orbitally-ordered AFM-insulating Pbnm phase back to PM-insulating R-3c phase at the same temperature at which the FM-metallic R-3c transforms to PM-insulating R-3c ,Fig.3b, it appears that a competing phase-separation evolves in $La_xMnO_{3\pm\delta}$ below $T_c$. Microscopically it is a competition between cooperative JT-distortion and DE. Continued increase of cooperative JTD above a certain limit would have triggered the R-3c to Pbnm transition but since it gets suppressed by the DE the nucleation and growth of the orbitally-ordered Pbnm phase gets effectively "arrested". This justifies the existence of kinetic-arrest like behavior with a characteristic temperature Tk ~235K, as seen in LT TEM studies. High-energy (200KV) e-beam irradiation of $La_xMnO_{3\pm\delta}$ somehow breaks the hybridization of l-b states of $e_g$-electrons and releases the growth of cooperative JTD causing nucleation and growth of the Pbnm phase. Drastic difference in the growth rates at 101K and 151K indicates that the probable value of $T^*$ should be ~140K. When the system is made "de-arrested" by e-beam trigger, say at 151K, it is still above $T^*$ and hence facing an energy-barrier. Due to energy-barrier growth rate will be slow and the latent heat gained by the transformed volume would not be sufficient enough to initiate further "de-arrested" and perpetual growth remains inhibited. But when triggering is done below $T^*$ the "de-arrested" supercooled volume does not see any energy-barrier hence the growth rate is much larger than that of at 151K. In this situation the rate of release of latent-heat will be large enough for further "de-arrest" and thus a chain reaction starts resulting in perpetual "de-arrest" and growth of the equilibrium phase. This is identical to devitrification and confirms that below ~235K $La_xMnO_{3\pm\delta}$ remain in kinetically arrested metastable state.

Thus we conclude that oxygen excess and self-doped $La_xMnO_{3\pm\delta}$ manganites under go the kinetic-arrest of the FOPT of R-3c to Pbnm at LT. The transformation is seen only when the R-3c phase, cooled sufficiently below 206K, is insitu triggered in TEM by focused e-beam irradiation. The observed kinetic-arrest is a result of competing phase-separation across the MI transition resulting in suppression of cooperative JTD by the DE.

**Acknowledgment**

Authors gratefully thank Dr. P. Chaddah, the Director and Prof. A. Gupta, the Center-Director of UGC-DAE-CSR, Indore for their encouragement and interest in the work. Dr. R. Rawat is sincerely acknowledged for his help in magneto transport measurements. Aga Shahee would like to acknowledge CSIR-India for financial support

.

**Figure caption**

**Fig.1** Typical [421] zone SAD patterns form R-3c phase of $La_xMnO_{3\pm\delta}$(x=1,0.9) (a) recorded immediately after the sample temperature got stabilized at 98K and (b) after aging the sample for ~12 hrs at 98K. (c) Electron-micrograph showing ~80nm nucleus formed after e-beam irradiation at 98K and (d) the corresponding SAD. Presence of extra spots, as indicated by arrows, shows the occurrence of a phase-transformation.

**Fig.2** (a) Electron-micrograph of the same region as shown in Fig.1c but recorded after ~1hour. The 80nm nucleus has now grown to a large region of ~1μm. (b) [001] zone SAD and (c) corresponding wide-angle CBED from the encircled part in Fig.2a of the well grown Pbnm phase. (d) [421] zone SAD of R-3c phase recorder after heating from 98K to RT.

**Fig.3** (a) Temperature dependent SADs showing vanishing of spot (indicated by arrows) intensity at higher temperatures. It indicates Pbnm to R-3c transformation. The temperature dependence of the spot intensity is shown in (b), right y-axis. The left y-axis of (b) shows amplitude of the e-beam deflection monitoring the FM to PM transition.

**Fig.4** Low-temperature XRD profiles of $La_xMnO_{3\pm\delta}$ (x=1, 0.9). It shows that no phase-transition is seen down to 79K. Inset shows the temperature variation of the profile of (220) peak of R-3c. Increase and decrease in FWHM across 235K can be seen.

**Fig.5** (a) Electron-micrograph showing formation of transformation nuclei after e-beam irradiation. This shows that the growth rate increase at lower temperatures. Series of EMs recorded after (b) 1 minute (c) 10 minutes (d) 20 minutes and (e) 40 minutes aging after the sample was e-beam irradiated at 101K.

**Fig.6** R-T data (left y-axis) and FWHM-vs-T data of XRD profiles of R-3c (220) peak (right y-axis) of $La_xMnO_{3\pm\delta}$ (x=1, 0.9). The R-T maxima corresponds to the MI-transition initiated by DE mediated FM transition.

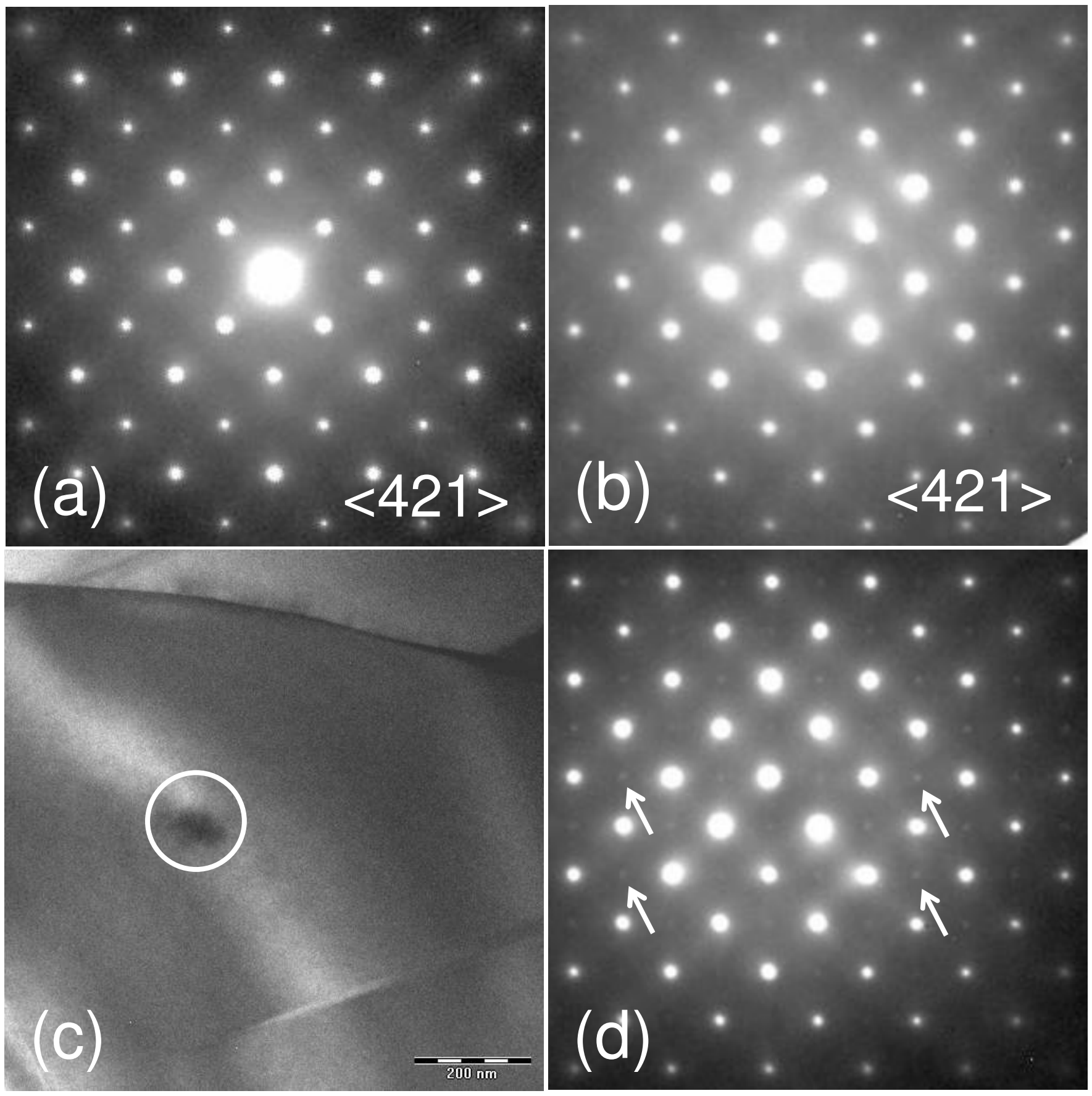

**Fig.1**

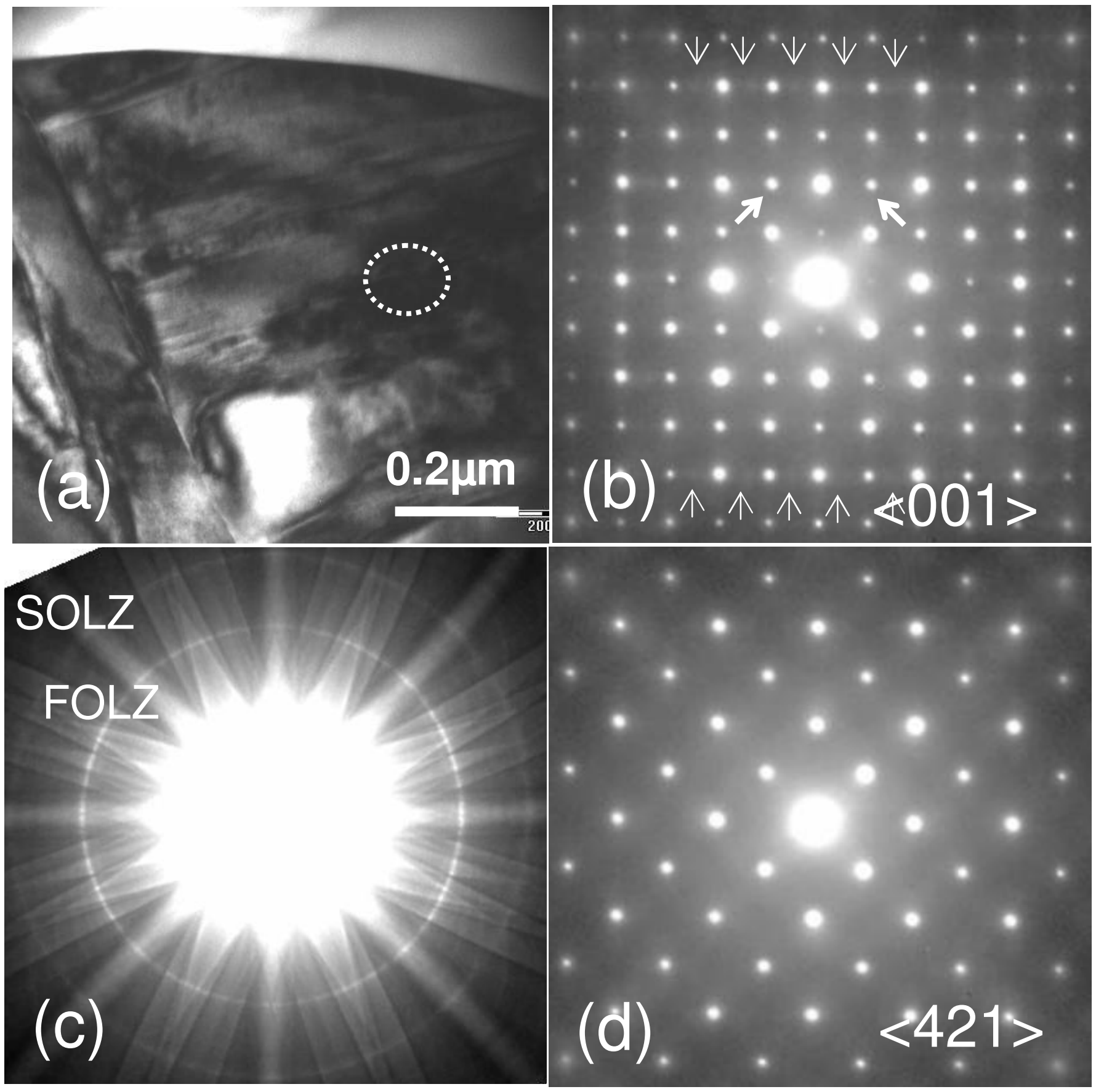

**Fig.2**

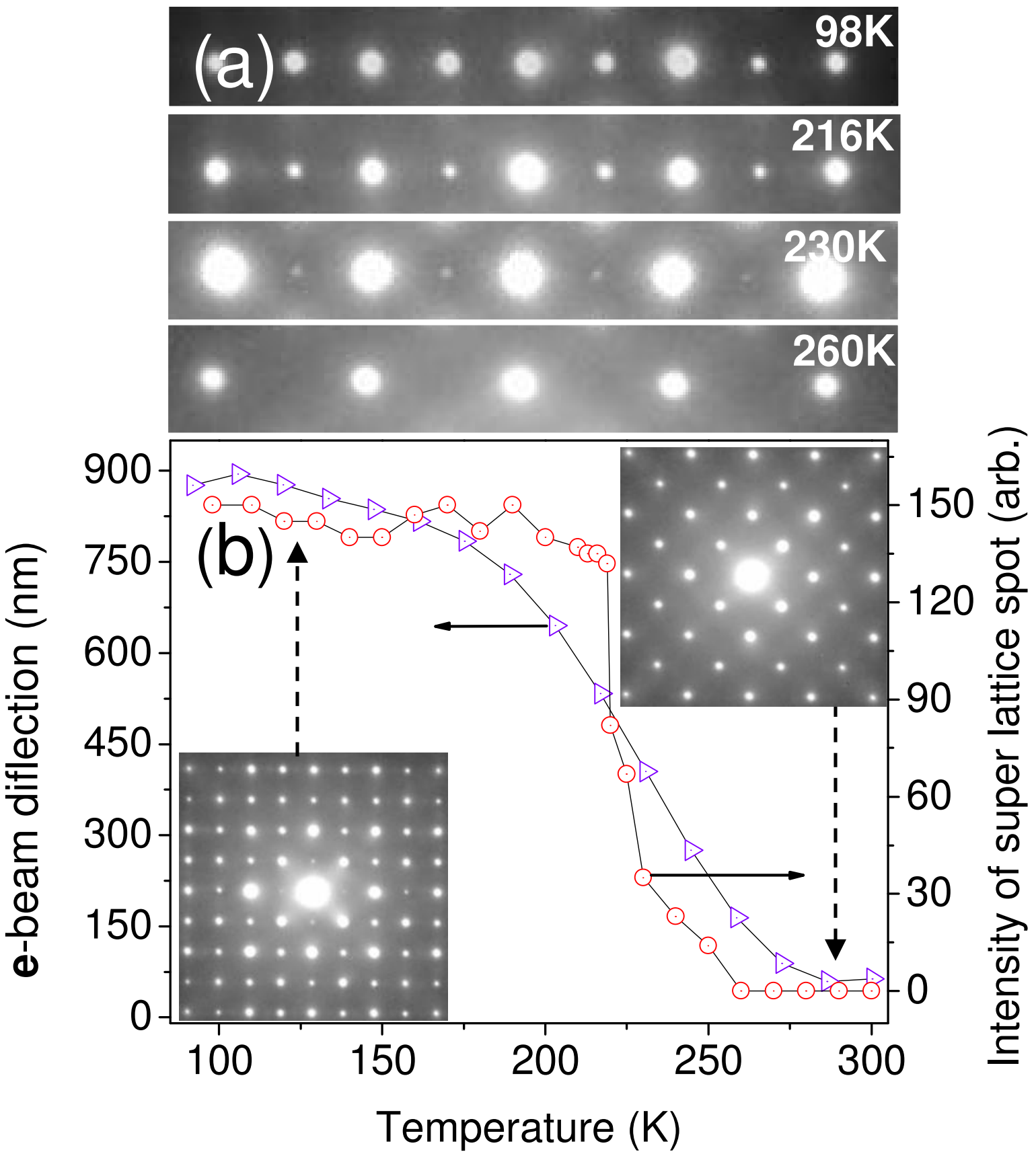

**Fig.3**

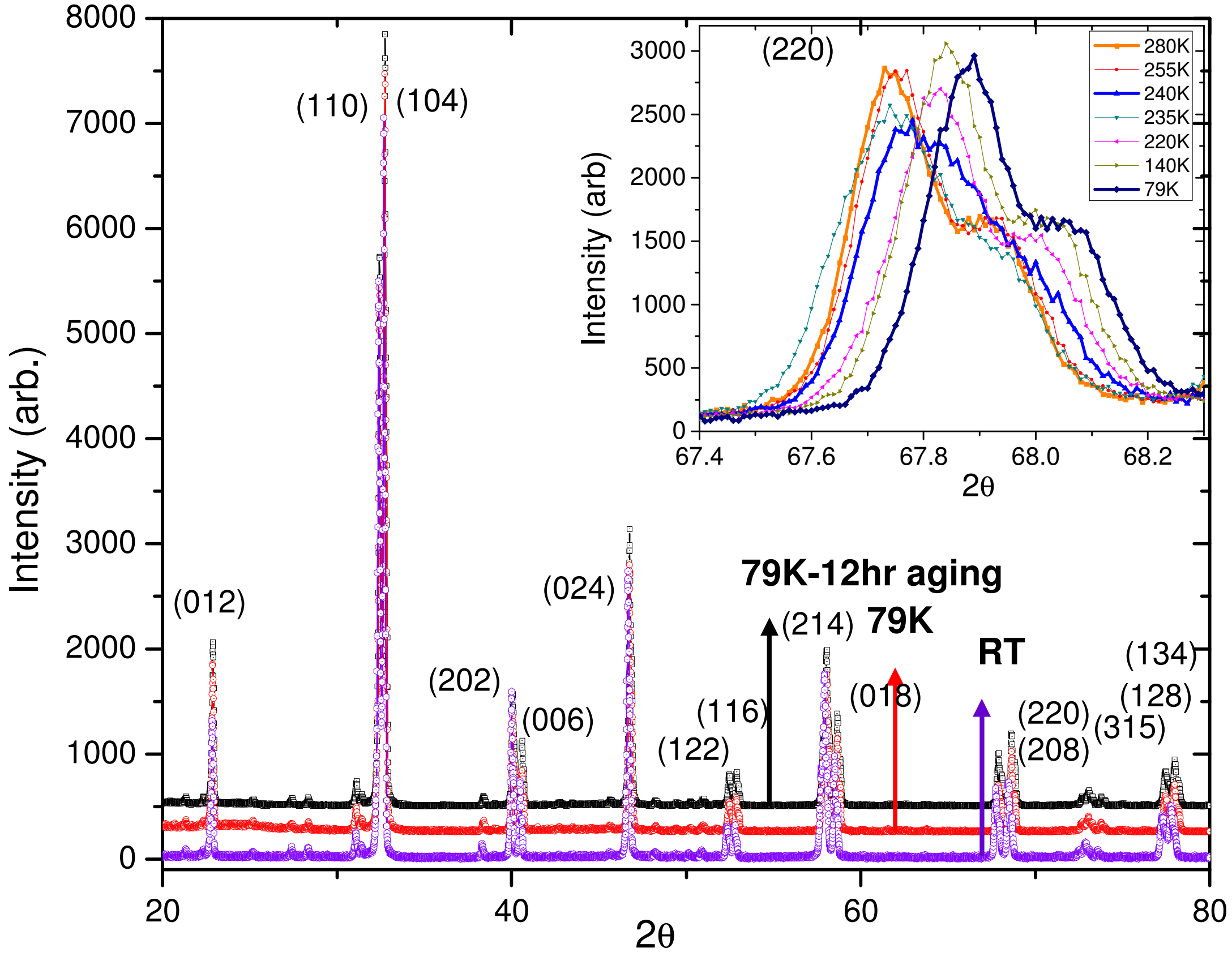

**Fig.4**

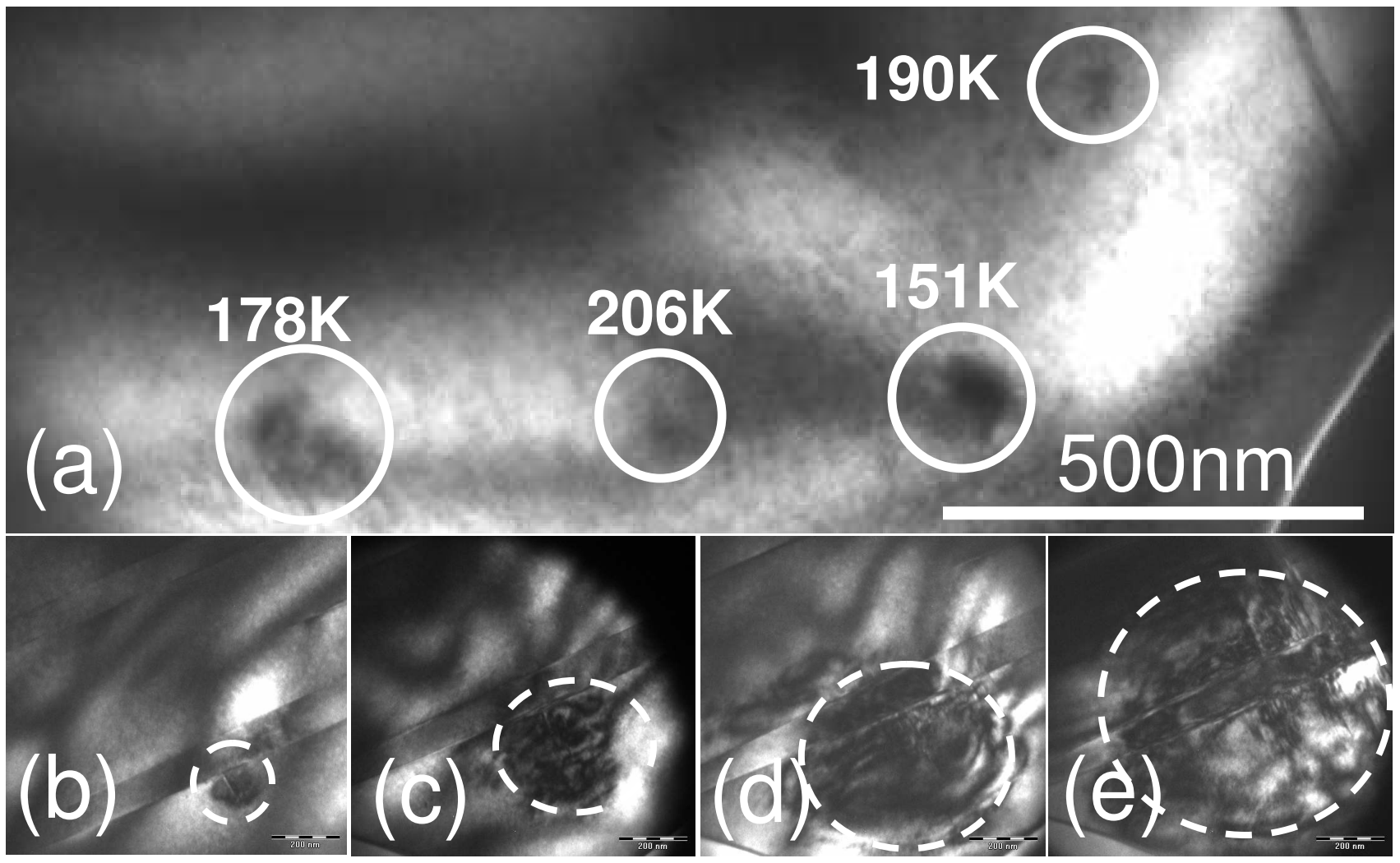

**Fig.5**

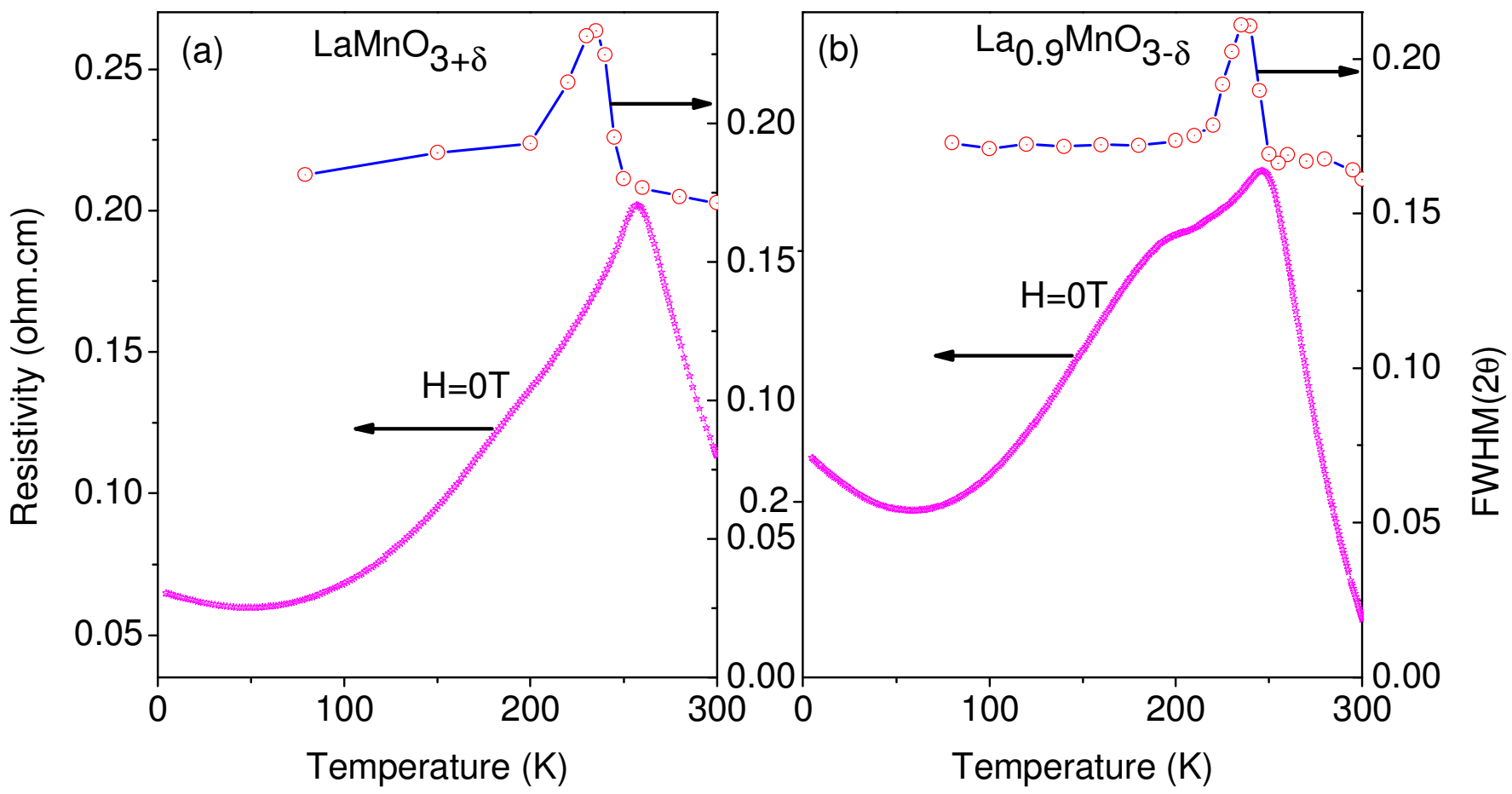

# Fig.6